\documentclass[preprint,12pt]{elsarticle}
\usepackage{amssymb}
\usepackage{amssymb}
\usepackage{amsmath}
\usepackage{color}
\usepackage{bm}
\usepackage{tabu}
\usepackage[colorlinks=true,allcolors=blue]{hyperref}%
\usepackage[top=1in, bottom=1in, left=1in, right=1in]{geometry}
\usepackage{graphicx}

\begin{document}

\begin{frontmatter}
\title{Inferring viscoplastic models from velocity fields: a physics-informed neural network approach}

\author[inst1,inst2]{Martin Lardy}

\affiliation[inst1]{adressine={Aix Marseille Univ, CNRS, Centrale Med, IRPHE, Turing Center for Living Systems, Marseille, France}}

\author[inst2]{Sham Tlili}
\author[inst1]{Simon Gsell}

\affiliation[inst2]{adressine={Aix-Marseille Univ, CNRS, IBDM, Turing Center for Living Systems, Marseille, France}}

\begin{abstract}

Fluid-like materials are ubiquitous, spanning from living biological tissues to geological formations, and across scales ranging from micrometers to kilometers. 
Inferring their rheological properties remains a major challenge, particularly when traditional rheometry fails to capture their complex, three-dimensional, and often heterogeneous behavior. 
This difficulty is exacerbated by system size, boundary conditions, and other material-specific physical, chemical, or thermal constraints.
In this work, we explore whether rheological laws can be inferred directly from flow observations.
We propose a physics-informed neural network (PINN) framework designed to learn constitutive viscoplastic laws from velocity field data alone. 
Our method uses a neural network to interpolate the velocity field, enabling the computation of velocity gradients via automatic differentiation. 
These gradients are used to estimate the residuals of the governing conservation laws, which implicitly depend on the unknown rheology. 
We jointly optimize both the constitutive model and the velocity field representation by minimizing the physical residuals and discrepancies from observed data.
We validate our approach on synthetic velocity fields generated from numerical simulations using Herschel–Bulkley, Carreau and Panastasiou models under various flow conditions. 
The algorithm reliably infers rheological parameters, even in the presence of significant noise. 
We analyze the dependence of inference performance on flow geometry and sampling, highlighting the importance of shear rate distribution in the dataset. 
Finally, we explore preliminary strategies for model-agnostic inference via embedded model selection, demonstrating the potential of PINNs for identifying the most suitable rheological law from candidate models.
This study illustrates how machine learning, and PINNs in particular, can enhance our ability to probe the rheology of complex fluids using velocity field data alone—paving the way for new approaches in computational rheology and material characterization.

\end{abstract}

\begin{graphicalabstract}
    \includegraphics[width=\linewidth]{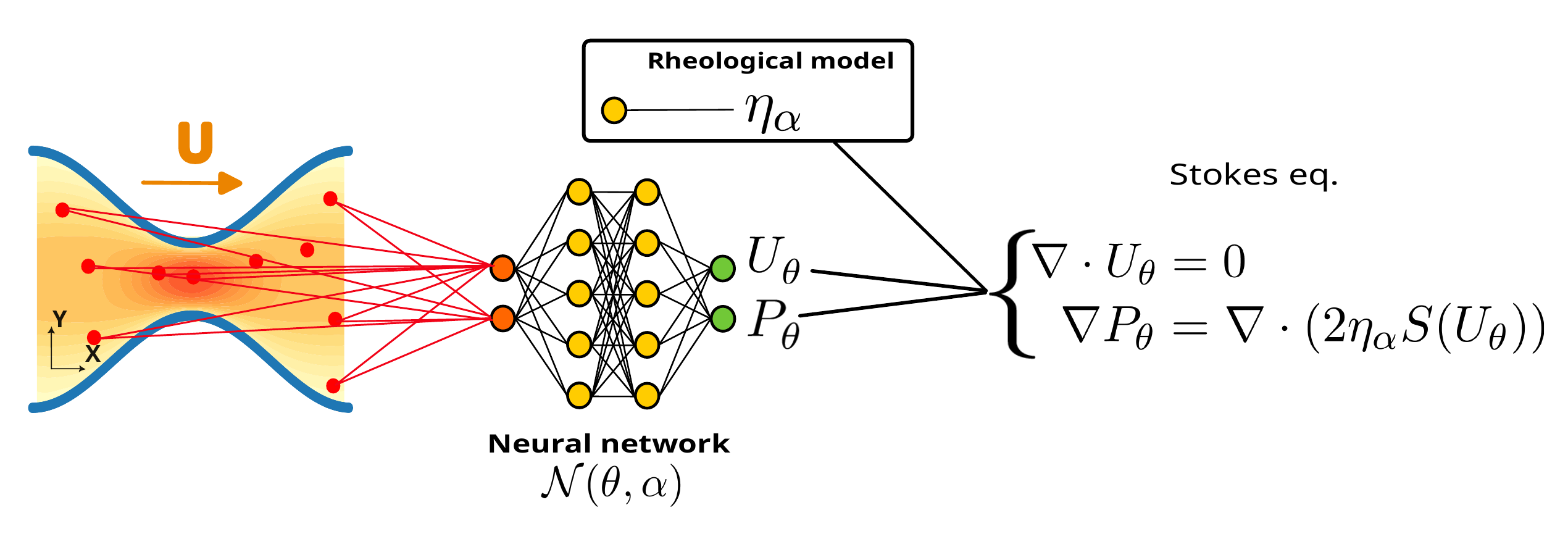}
\end{graphicalabstract}

\newpage
\
\end{frontmatter}
\section{Introduction}

Many natural and industrial materials exhibit complex rheological behavior, where deformation responds nonlinearly to the stresses driving it. 
Accurately modeling these rheological properties is crucial for understanding and predicting how materials deform—ranging from micrometer-scale processes such as tissue morphogenesis \cite{lenne22}, to kilometer-scale geodynamic flows in the Earth's mantle \cite{davaille18, tommasi2000viscoplastic}.

Traditional rheometry involves bringing a sample to the laboratory and subjecting it to controlled stress or strain to characterize its constitutive behavior. 
This technique has been key to understand the mechanics of complex materials such as granular media \cite{jop06} and dense suspensions \cite{guazzelli18, etcheverry23}. 
However, this approach also has several limitations. 
First, it typically provides only one-dimensional, local measurements that may not generalize to describe large-scale or spatially heterogeneous phenomena involving non-local effects or anisotropy. 
Second, many materials are difficult or impossible to study using standard rheometers—due to size constraints (from biological tissues to planetary layers), interfacial tribology (e.g., soft materials that do not adhere well to solid surfaces), or extreme physical conditions (e.g., high temperatures in lava, low temperatures in snow).

An alternative to conventional rheometry is to infer rheological properties directly from the way a material flows in situ—by observing its deformation in complex, natural, or engineered configurations. 
Across scales, material deformation is governed by the same fundamental conservation laws for mass and momentum. 
This raises the question: Can we invert these equations using experimental flow data to recover the constitutive law of the material?

Machine learning techniques have recently been developed to tackle this challenge by identifying unknown terms in partial differential equations (PDEs) that describe physical systems \cite{Brunton_2016, bongard2007automated, rudy2017data, kaheman2020sindy, chen2021physics, colen2021machine}. 
In fluid mechanics, such methods have been used to infer hidden features such as boundary conditions, external forces, or material properties, using approaches ranging from Bayesian inference \cite{christopher2018parameter, kontogiannis2024learning} to neural networks \cite{fan2020solving, xu21, brunton2020machine}.

Among these, Physics-Informed Neural Networks (PINNs) have emerged as a promising framework for solving forward and inverse problems in physics and engineering \cite{raissi2019physics, lagaris1998artificial}. 
In PINNs, physical variables such as velocity or stress are approximated using neural networks, and training is driven by minimizing a loss function that combines data fitting with residuals from governing PDEs. 
Automatic differentiation is used to compute spatial and temporal derivatives efficiently \cite{baydin18}, enabling the enforcement of physical constraints during learning. 
This framework naturally accommodates inverse problems, allowing unknown parameters—including those in constitutive laws—to be inferred as part of the optimization. 
PINNs have been successfully applied to a variety of fluid mechanics problems \cite{cai2021physics}, including the recovery of missing flow data \cite{raissi20, lou2021physics}, and inference of power-law \cite{zhai2023deep} or viscoelastic \cite{thakur2024viscoelasticnet} properties from velocity fields, as well as viscoelastoplastic laws from rheometry data \cite{mahmoudabadbozchelou2021rheology, mahmoudabadbozchelou22}.

In this work, we address the problem of inferring viscoplastic constitutive laws directly from flow velocity data using a PINN-based approach. 
Our algorithm employs a single neural network to represent both the velocity and pressure fields. 
We apply it to synthetic velocity fields generated from simulations of steady flows in confined geometries, inspired by microfluidic setups used to probe the rheology of foams \cite{cheddadi11}, emulsions \cite{golovkova20}, and living tissues \cite{tlili20, tlili_microfluidic_2022}. 
These flows are computed using established viscoplastic models, including Herschel–Bulkley, Carreau, and Papanastasiou models.

We first demonstrate that, given prior knowledge of the correct constitutive model, the algorithm can accurately infer its parameters using only a single velocity snapshot—also reconstructing the corresponding pressure field. 
Importantly, the method remains robust in the presence of artificial noise, which is promising for application to experimental data. 
We systematically analyze how inference performance depends on the structure of the input data, including the number and spatial distribution of velocity measurements. 
Our results highlight that a broad distribution of shear rates in the sampled data is critical for successful parameter recovery, and we propose a shear-rate-biased sampling strategy to improve performance.
Finally, we present a preliminary strategy for model selection, enabling our algorithm to identify the most appropriate rheological model from a set of candidates before optimizing the associated parameters.
Altogether, we believe our work lays a foundation for automated rheology inference from velocity fields, offering a data-driven alternative to classical rheometry that may later be applied to complex or experimentally constrained systems.

\section{Rheology inference approach}

\subsection{Inverse problem formulation}
We focus on two-dimensional incompressible fluid flows in creeping regime, governed by the Stokes equations:
\begin{subequations}
\begin{align}
    &\nabla \cdot \bm{U} = 0, \\
    &\nabla P = \nabla \cdot [ 2 \eta S(\bm{U}) ],
\end{align}
\label{eqn:Stokes}
\end{subequations}
with $\bm{U} = (u,v)$ the velocity field, $P$ the pressure and $S(\bm{U}) = (\nabla \bm{U} + (\nabla \bm{U})^T)/2$ the shear-rate tensor.
For a viscoplastic fluid, the effective viscosity $\eta$ depends on the local shear rate $\dot{\gamma} = \sqrt{2S:S}$.
In the following, we aim at inverting equations (\ref{eqn:Stokes}) for a prescribed velocity field $\bm{U}$, i.e. finding the viscosity function $\eta$ and the pressure field $P$ such that equations (\ref{eqn:Stokes}a-b) are satisfied.
In order to avoid the trivial solution $\eta=0$ and $P=const.$, we introduce a characteristic viscosity $\eta_c\ne 0$ such that $\eta=\eta_c \eta^*$, with $\eta^*$ a non-dimensional viscosity.
The other flow quantities are also made non-dimensional as follows: 
\begin{equation}
S^*=\dfrac{S}{\dot{\gamma_c}},~~ U^*=\dfrac{U}{U_0},~~ \bm{X}^*=\dfrac{\bm{X}}{D},~~ P^*=\dfrac{P}{\eta_c\dot{\gamma_c}}, 
\end{equation}
with $\bm{X} = (x,y)$ the spatial coordinates, $U_0$ a characteristic velocity (for example the mean flow velocity), $D$ a characteristic length and $\dot{\gamma_c}= U_0 / D$ the characteristic shear rate.
The non-dimensional form of equations (\ref{eqn:Stokes}a-b) reads
\begin{subequations}
\begin{align}
        &\nabla \cdot \bm{U}^* = 0, \\
        &\nabla P^* = \nabla\cdot (2 \eta^* S^*).
\end{align}
\label{eqn:Stokes*}
\end{subequations}
In the following, we introduce three different rheological models for $\eta(\dot{\gamma_c})$, namely the Herschel-Bulkley, Carreau and Papanoustasiou \cite{papanastasiou1987flows} models.

\subsection{Viscoplastic models}

\begin{table}[]
    \centering

\tabulinesep=1.2mm
\begin{tabu}{| c | c | c | c | c |}
    \hline
    & \small  \# of parameters & \small $\alpha_0$ & \small $\alpha_1$ & \small $\alpha_2$ \\
    \hline
    \small Herschel-Bulkley (\ref{eqn:HB*}) & \small 2 & $\dfrac{\sigma_0}{k\dot{\gamma}_c^{n}}$ & \small n & \small - \\
    \hline
    \small Carreau (\ref{eqn:carreau*}) & \small 3 & \small $\dfrac{\eta_\infty}{\eta_0}$ & \small $\lambda \dot{\gamma}_c$ & \small n \\
    \hline
    \small Papanastasiou (\ref{eqn:papa*}) & \small 2 & $\frac{\tau_0}{\mu_0}$ & \small $m \dot{\gamma}_c$ & \small - \\
    \hline
\end{tabu}
\caption{Summary of the rheological models considered in this work. The non-dimensional rheological parameters $\alpha_i$ are obtained by choosing the characteristic viscosities $\eta_c=k\Dot{\gamma}_c$ (Herschel-Bulkley), $\eta_c=\eta_0$ (Carreau) and $\eta_c=\mu_0$ (Papanastasiou).}
\label{tab:rheology} 
\end{table}

We present here the three rheological models considered in our work and their non-dimensional form. 
The models are summarized in table \ref{tab:rheology}.

\textbf{Herschel-Bulkley model}. The Herschel-Bulkley model accounts for shear-thinning/thickening and yield-stress behaviors of the fluid. 
This model is popular to characterize viscoplastic properties, and has been successfully used to describe the steady rheology of soft glassy and jammed materials or polymeric gels \cite{bonn2017yield}.
The viscosity function reads 
\begin{equation}
    \eta_{HB} =\frac{\sigma_0}{\Dot{\gamma}}+k\dot{\gamma}^{n-1},
    \label{eqn:HB}
\end{equation}
with $\sigma_0$ the yield stress, $k$ the fluid consistency and $n$ the power-law index. 
We choose $\eta_{c}=k \dot{\gamma}_c^{n-1}$ as the characteristic viscosity, assuming $k \ne 0$.
The non-dimensional viscosity becomes 
\begin{equation}
  \centering
  \eta^*_{HB}=\frac{\alpha_0}{\dot{\gamma}^*}+(\dot{\gamma}^*)^{\alpha_1-1},
  \label{eqn:HB*}
\end{equation}
with $\alpha_0=\dfrac{\sigma_0}{k\dot{\gamma}_c^{n}}$ and $\alpha_1=n$ the two non-dimensional rheological parameters to be inferred.
Note that $\alpha_0$ corresponds to the standard definition of the Bingham number for yield-stress fluids.

\textbf{Carreau model}. The Carreau model only accounts for shear-thinning / thickening effects, but the power-law behavior of the fluid is localized within a certain shear-rate range.
Outside of the power-law region, the fluid recovers a Newtonian behavior.
The viscosity reads
\begin{equation}
    \eta_C =\eta_\infty+(\eta_0-\eta_\infty)(1+(\lambda\dot{\gamma})^2)^{\frac{n-1}{2}},
    \label{eqn:carreau}
\end{equation}
where $\eta_0$ and $\eta_\infty$ are the limit viscosities at zero and infinite shear rates, respectively, $\lambda$ is the characteristic time and $n$ is the power-law index.
The Carreau model thus involves one extra parameter compared to the Herschel-Bulkley model.
We choose $\eta_{c}=\eta_0$ as the characteristic viscosity, leading to the following non-dimensional viscosity law:
\begin{equation}
    \eta^*_{C} = \alpha_0+(1-\alpha_0)(1+(\alpha_1 \dot{\gamma}^*)^2)^{\frac{\alpha_2-1}{2}},
    \label{eqn:carreau*}
\end{equation}
with $\alpha_0= \eta_\infty / \eta_0$, $\alpha_1= \lambda  \dot{\gamma_c}$ and $\alpha_2= n$ the three non-dimensional parameters to be inferred.

\textbf{Papanastasiou model}. This model is a regularized version of the Bingham model (i.e. Herschel-Bulkley model with $n=1$), accounting for yield-stress behavior but avoiding the divergence of the viscosity at zero shear rate.
The viscosity model reads
\begin{equation}
    \eta_P =\left( 1-e^{-m\dot{\gamma}} \right) \frac{\tau_0}{\dot{\gamma}}+\mu_0,
    \label{eqn:papa}
\end{equation}
with $\mu_0$ the limit viscosity at zero shear rate, $\tau_0$ the yield stress and $m$ the characteristic time.
We choose $\eta_c=\mu_0$, giving the following non-dimensional viscosity:
\begin{equation}
    \eta_P^* =(1-e^{-\alpha_1\dot{\gamma}^*})\frac{\alpha_0}{\dot{\gamma}^*}+1,
    \label{eqn:papa*}
\end{equation}
with $\alpha_0=\dfrac{\tau_0}{\mu_0 \dot{\gamma_c}}$ and $\alpha_1= m \dot{\gamma_c}$ the non-dimensional rheological parameters to be inferred.

\subsection{Rheological inference approach}

\begin{figure}[t]
    \centering
    \includegraphics[width=1.05\linewidth,keepaspectratio]{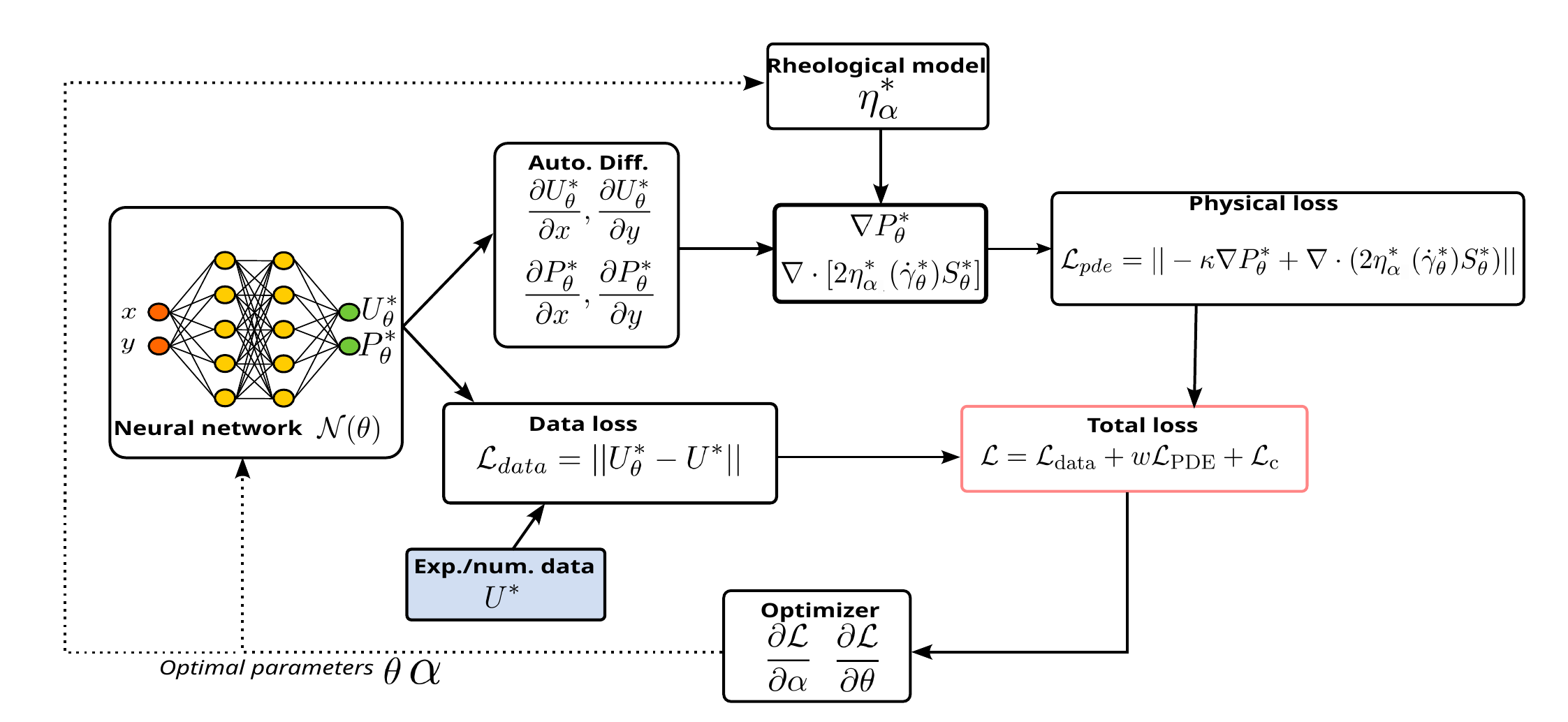}
    \vskip-0.5em
    \caption{Overview of our PINN rheology inference algorithm.
    We use a neural network $\mathcal{N}_\theta$ to represent the flow quantities $\bm{U_\theta^*}$ and $P_\theta^*$.
    The distance between the output velocity $\bm{U_\theta^*}$ and the velocity data $\bm{U^*}$ is quantified by the data loss $\mathcal{L}_{data}$ (\ref{eqn:Ldata}).
    We then use automatic differentiation to determine the spatial gradients of $\bm{U_\theta^*}$ and $P_\theta^*$, that are then combined with the rheological model $\eta_\alpha^*$ to estimate the physical loss $\mathcal{L}_{phy}$ (\ref{eqn:Lphy}).
    The total loss consists of the weighted sum of $\mathcal{L}_{data}$ and $\mathcal{L}_{phy}$, and also includes an additional loss $\mathcal{L}_{\text{c}}$ (\ref{eqn:Lc}) enforcing the positivity of the rheological parameters $\alpha$.
    The rheology inference is performed by minimizing the total loss over the physical/rheological parameters $\alpha$ and interpolation parameters $\theta$.
    \label{fig:scheme_PINNs}
    }
\end{figure}

Our inference algorithm is summarized in figure \ref{fig:scheme_PINNs}.
We aim at inverting equations (\ref{eqn:Stokes*}a-b) by determining the viscosity function $\eta^*_\alpha$, with $\alpha=\{ \alpha_0,\alpha_1,...\}$ the parameters of one of the three rheological models above.
We perform the inverse problem from a set of velocity samples $\bm{U_n} = (u_n, v_n)$, $n\in\{1,...,N\}$ with $N$ the total number of samples.
Each sample is normalized as 
\begin{equation}
    \bm{U^*_n}=\frac{(u_n-\Bar{u},v_n-\Bar{v})}{U_0},
    \label{eqn:Unorm}
\end{equation}
where $\Bar{u}$ and $\Bar{v}$ are the mean values of the $u$ and $v$ fields.
The position $\bm{X_n}$ of each velocity sample is also normalized as $\bm{X_n^*}=\bm{X_n}/D$.

Assuming that the data $\bm{U^*_n}$ might be noisy and sparse, we first perform an interpolation of $\bm{U^*_n}$ with a field $\bm{U^*_\theta}(\bm{X}^*)$, where $\theta$ is a set of interpolation parameters.
In the PINN approach, we use a neural network $\mathcal{N}_\theta$ to represent the function $\bm{U^*_\theta}(\bm{X}^*)$, with $\theta$ the bias and weights of the network.
In practice, the output of the network is not directly $\bm{U^*_\theta}$ but the stream function $\psi_\theta$; the velocity is then reconstructed as $u_\theta = \partial \psi_\theta / \partial y^*$ and $v_\theta = -\partial \psi_\theta / \partial x^*$.
This allows us to enforce the incompressibility condition (\ref{eqn:Stokes*}a) for the velocity field $\bm{U^*_\theta}$. In a compressible scenario, then the output would directly be $u_\theta$ and $v_\theta$.
We also include the unknown pressure $P^*_\theta$ as an output of the same network $\mathcal{N}_\theta$.
The neural network used here is a fully connected feedforward network composed of 4 hidden layers with 20 neurons each, with a $\tanh$ activation function.

In order to train the network $\mathcal{N}_\theta$, we first introduce the data loss $\mathcal{L}_{data}$ which quantifies the distance between the interpolated velocity field and the data $\bm{U_n}$,
\begin{equation}
    \mathcal{L}_{data}= \lVert \bm{U^*_\theta} - \bm{U^*_n} \rVert,
    \label{eqn:Ldata}
\end{equation}
with $\lVert . \rVert$ the 2-norm squared ($\lVert . \rVert_2^2=\sum^N (.)^2$), with N the number of training points. Assuming a guess rheological model $\eta_\alpha^*$, we then define the physical loss $\mathcal{L}_{phy}$ based on the momentum balance residual (\ref{eqn:Stokes*}b),
\begin{equation}
    \mathcal{L}_{phy} = \lVert -\kappa\nabla P^*_\theta+ \nabla\cdot (2 \eta^*_{\alpha} S^*_\theta) \rVert,
    \label{eqn:Lphy}
\end{equation}
where $\nabla P^*_\theta$ and $S^*_\theta$ are determined using automatic differentiation of the network $\mathcal{N}_\theta$.
Even though not necessary, in the following we choose to calculate $\mathcal{L}_{phy}$ at the same colocation points as for $\mathcal{L}_{data}$.
In (\ref{eqn:Lphy}) the pressure field can be rescaled by any arbitrary prefactor $\kappa$.
As the pressure field magnitude depends on the characteristic viscosity $\eta_c$, we have found that the inference is improved when rescaling the pressure using $\kappa \propto 1 / \eta_c$.
In practice, we use $\kappa = \dot{\gamma_c}^{n-1}$ for the Herschel-Bulkley model and $\kappa = 1$ for the Carreau and Papanastasiou models.
In the case of the Herschel-Bulkley model, this allows to encode the rescaling of the pressure by the power-law index $n$, so that the neural network does not have to learn this rescaling each time the rheological model is updated.
In order to enforce the positivity of the rheological parameters, we finally introduce a positivity loss 
\begin{equation}
    \mathcal{L}_p=\sum_i \max(0,-\alpha_i).
    \label{eqn:Lc}
\end{equation}

The total loss defining our regression problem finally reads 
\begin{equation}
  \mathcal{L}=\mathcal{L}_{data}+W\mathcal{L}_{pde}+\mathcal{L}_p,
  \label{eqn:total_loss}
\end{equation}
where $W$ is a constant weight used to balance the data and physical losses.
In the following, we automatically fix the value of $W$  at the beginning of each simulation, by computing the initial ratio between $\mathcal{L}_{data}$ and $\mathcal{L}_{phy}$ such that $\lfloor \log_{10}(\frac{\mathcal{L}_{data}}{\mathcal{L}_{phy}})\rfloor+1=w$ and $W=10^{w-1}$.
During the rheology inference process, we aim to minimize the total loss $\mathcal{L}$ by performing a descent along gradients $\partial \mathcal{L}/\partial \theta$ and $\partial \mathcal{L}/\partial \alpha$ with respect to the interpolation and physical/rheological parameters, respectively.
We perform this gradient descent using the ADAM algorithm \cite{kingma2014adam}, with a learning rate equal to $0.0025$.
When the optimal parameters $\alpha$ and $\theta$ are found, the algorithm provides a rheological model $\eta_\alpha$ minimizing the physical residual for a velocity field $\bm{U_\theta}$ as close as possible to the velocity data $\bm{U_n}$. All the Python scripts used in this work are based on the Pytorch library and are available at \href{https://github.com/martinLARD/Pinn.git}{https://github.com/martinLARD/Pinn.git} \cite{lardygit}.

\subsection{Benchmarking approach}

We test our rheology inference approach using a series of non-Newtonian flows generated by numerical simulations.
The simulations are based on the lattice-Boltzmann and immersed boundary methods \cite{gsell2021lattice,gsell2021direct}.
We perform simulations in different configurations similar to those classically used experimentally to characterize foams \cite{cheddadi11}, emulsions \cite{golovkova20}, or cellular tissues \cite{tlili20,tlili_microfluidic_2022}, such as a straight channel, different types of non-uniform channels, and a channel including a circular obstacle. 
The simulations are carried out for different viscoplastic models and various values of the associated parameters.
Our lattice-Boltzmann method is based on a generalized-Newtonian approach, i.e. viscoplastic behaviors are purely modeled by locally varying the effective fluid viscosity, as described in detail in \cite{gsell2021lattice}.
The velocity fields resulting from the simulations are sampled uniformly or with a bias based on the shear-rate distribution (see Section \ref{sec:dfferent_geo}). 
For each inference run, the sampling points remain fixed during the calculation.
We quantify the inference accuracy by calculating the relative log-viscosity error between the true viscosity function $\eta^*$ and the inferred viscosity $\eta_\alpha^*$,
\begin{equation}
    E_\eta=\frac{1}{M}\sum_{i=0}^M \frac{|\log(\eta_i^*)-\log(\eta^*_{\alpha,i}(\dot{\gamma}^*))|}{|\log(\eta_i^*) |},
    \label{eqn:error}
    \end{equation}
where $M$ is the number of $\eta^*$ points available.

\section{Inference of rheological parameters}

\subsection{Overview of the inference process}

\begin{figure}[t]
     \centering
        \includegraphics[width=\textwidth]{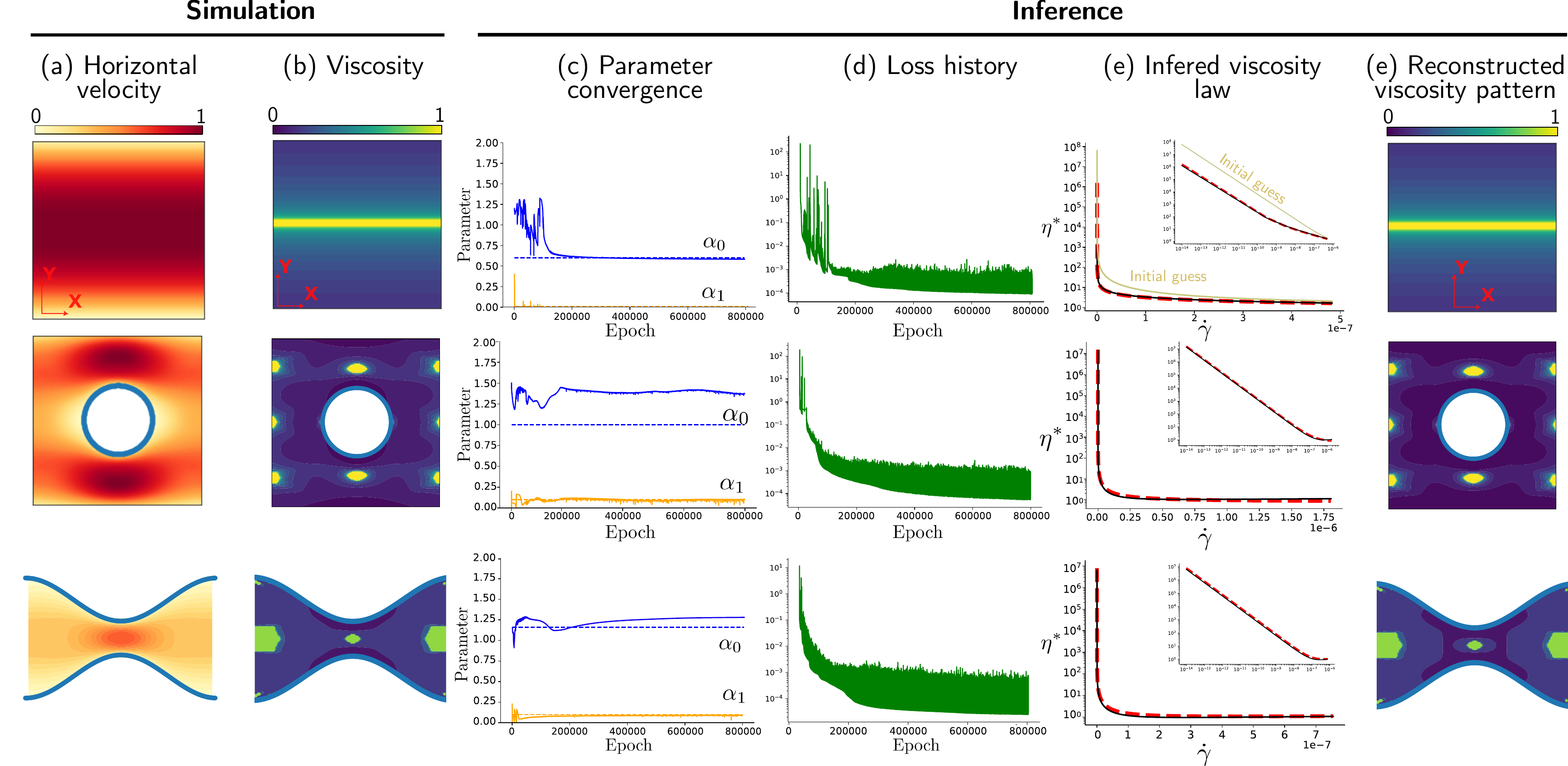}
    \caption{Overview of the inference process by infering Herschel-Bulkley parameters in different configurations.
    (a,b) We perform simulations of Herschel-Bulkley flows with different values of the parameters $\alpha_0$ and $\alpha_1$ in a uniform channel (top), a channel obstructed by a circular obstacle (center) and a non-uniform channel (bottom). The velocity field (column (a)) is sampled as the input data for the inference algorithm, while the viscosity field (column (b)) is unknown during the inference.
    (c,d) As the total loss $\mathcal{L}$ decreases during learning, the algorithm converges to inferred values for $\alpha_0$ and $\alpha_1$ (solid lines) that are close the true values (dashed lines).
    (e,f) The viscosity functions reconstructed using the inferred values of $\alpha_0$ and $\alpha_1$ match the true viscosity functions, both as a function of the shear rate (column (e)) or in space (column (f)).}

    \label{fig:overview}
\end{figure}

Figure \ref{fig:overview} shows a series of inference examples applied to Herschel-Bulkley flows in a uniform channel, a channel obstructed by a circular obstacle and a non-uniform channel.
In all cases, we perform numerical simulations providing velocity fields (figure \ref{fig:overview}a) from which we uniformly sample a total of 800 velocity data points provided to the inference alogrithm.
We will further discuss the effect of the colocation points in section 3.1.1 and section 4.
Each configuration leads to a distinct viscosity pattern (figure \ref{fig:overview}b) that needs to be infered by the algorithm.
We fix the number of epochs/iterations to $6\times 10^5$.

In all cases, the learned rheological parameters $\alpha_0$ and $\alpha_1$ converge towards their true values (figure \ref{fig:overview}c) as the total loss decreases during the training process (figure \ref{fig:overview}d).
However, the accuracy of the parameter inference depends on the relative influence of each specific parameter on the resulting viscosity field. 
We thus further evaluate the inference quality by reconstructing the viscosity law $\eta^*(\dot{\gamma})$ and the viscosity field $\eta(\bm{x})$ from the infered parameters (figures \ref{fig:overview}e,f).
In all cases, the reconstructed viscosity matches with the true one, even when small errors are observed on the rheological parameters (e.g. $\alpha_0$ for the flow past the obstacle).

\begin{figure}[!t]
    \centering
    \includegraphics[width=\textwidth]{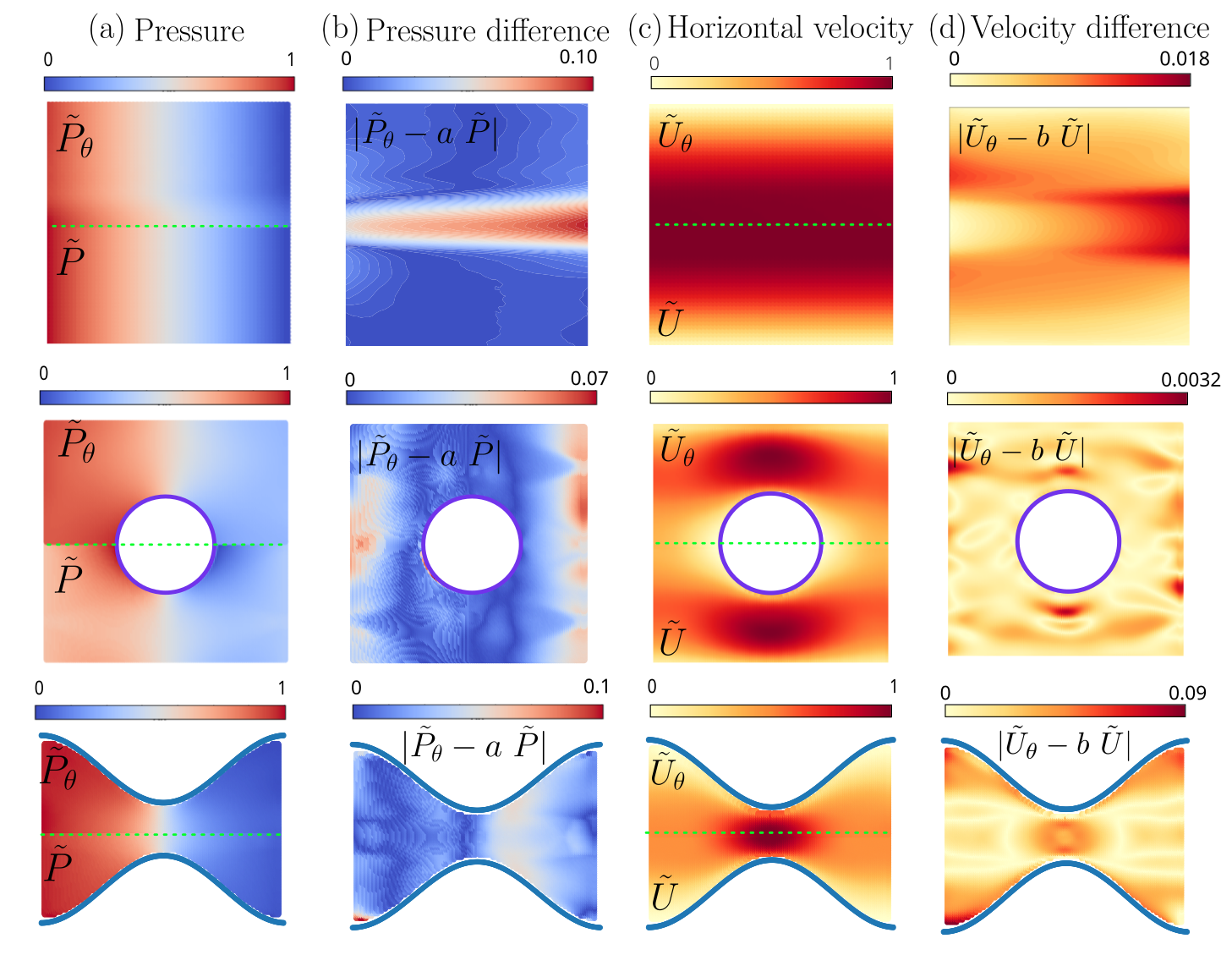}
    \caption{Velocity and pressure fields at the output of the neural network, for the three examples shown in figure (\ref{fig:overview}). 
    We introduce the notation $\tilde{A}=\frac{A-\min(A)}{\max(A)-\min(A)}$, for $A$ a given spatial field. 
    The figure shows Herschel-Bulkley examples with parameters $(\alpha_0,\alpha_1)$, from top to bottom, equal to $(1.2,0.1)$, $(1,0.1)$ and $(1,0.1)$.
    (a) Pressure fields. The upper part of the plot shows the output of the neural network ($\bar{P}^*_\theta$); the lower part shows the pressure obtained from numerical simulation ($P$) both normalized.
    (b) Absolute difference between both pressure fields with a factor $a$ before the normalized pressure tuned to minimize the absolute difference integrated on the entire channel. This parameter is from top to bottom equal to 0.94, 1 and 0.93.
    (c) Horizontal velocity fields. The upper part of the plot shows the output of the neural network; the lower part shows the velocity obtained from numerical simulation.
    (d) Absolute difference between both velocity fields, with b from top to bottom equal to 1.07, 0.98 and 0.96.
    }
    \label{fig:P_U_field}
\end{figure}

We plot the output fields of the neural network and compare them to the true fields in figure \ref{fig:P_U_field}.
The overall pressure pattern is well predicted by the PINN algorithm, even though pressure boundary conditions are not provided during the inference process.
Some discrepencies are however visible between the original and inferred pressure fields, especially near solid boundaries in the flow past the cylinder and in the wavy channel.
In the uniform channel case, pressure error mostly occurs in the central plug region, where the fluid is unyielded and the effective viscosity is very high (figure \ref{fig:overview}b).
Overall, these pressure errors however remain smaller than 10\% during the inference (figure \ref{fig:P_U_field}b).
As expected, discrepancies are even smaller for the velocity fields (figure \ref{fig:P_U_field}c,d), due to the minimization of $\mathcal{L}_{data}$.
Again, largest errors are observed close to solid boundaries and within the unyielded regions, but overall the velocity error remains smaller than 10\% for all configurations.

\subsubsection{Effect of colocation points}

\begin{figure}[t]
    \centering
    \includegraphics[width=\textwidth]{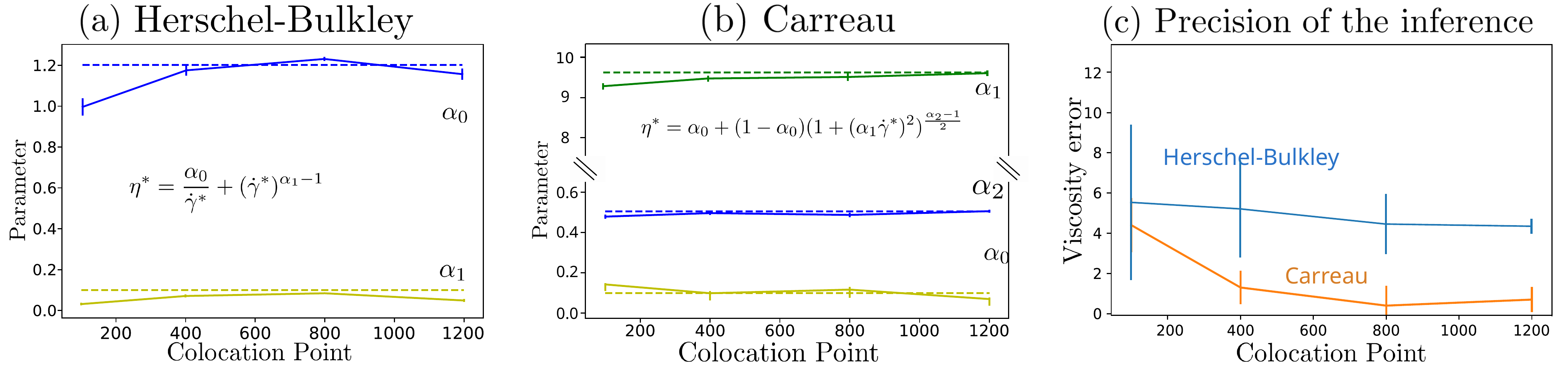}
    \caption{Impact of the number of colocation points on the inference accuracy.
    A series of 20 inferences are performed during 800 000 epochs, each conducted with either 100, 400, 800, or 1200 collocation points, for (a) the Herschel-Bulkley and (b) Carreau flows in a uniform channel.
    (a,b) Mean values of the infered rheological parameters. The error bars indicate the standard error of the mean.
    (c) Mean viscosity error (\ref{eqn:error}) for both rheological models. Error bars indicate the error propagated from the rheological parameter values.
    }
    \label{fig:coloc_point}
\end{figure}

We quantify the impact of the number of colocation points on the inference accuracy in figure \ref{fig:coloc_point}.
We perform a series of 20 inferences with either 100, 400, 800, or 1200 collocation points and $8 \times 10^5$ epochs, for (a) the Herschel-Bulkley and (b) Carreau flows in a uniform channel.
We note that the inference accuracy is generally higher for the Carreau model.
For both rheological models, the inference accuracy generally increases with the number of points, which can be noted both from the values of the inferred parameters (figure \ref{fig:coloc_point}a,b) and from the overall viscosity law error (figure \ref{fig:coloc_point}c).
Interestingly, the inference variability also drastically decreases with the number of colocation points.
This is particularly visible from the viscosity error in the case of the Herschel-Bulkley rheology (figure \ref{fig:coloc_point}c).
This can be interpreted as a decrease of the inference uncertainty as the amount of input information is increased.
However, for 800 colocation points and more, the inference accuracy does not significantly decrease with the number of points anymore; it even starts to increase again in the case of the Carreau model.
Similar detrimental effect of a too large amount of colocation points has also been reported in previous works on PINNs \cite{chen2025global}.

\subsection{Noise robustness}

\begin{figure}[t!]
     \centering
        \includegraphics[width=\textwidth]{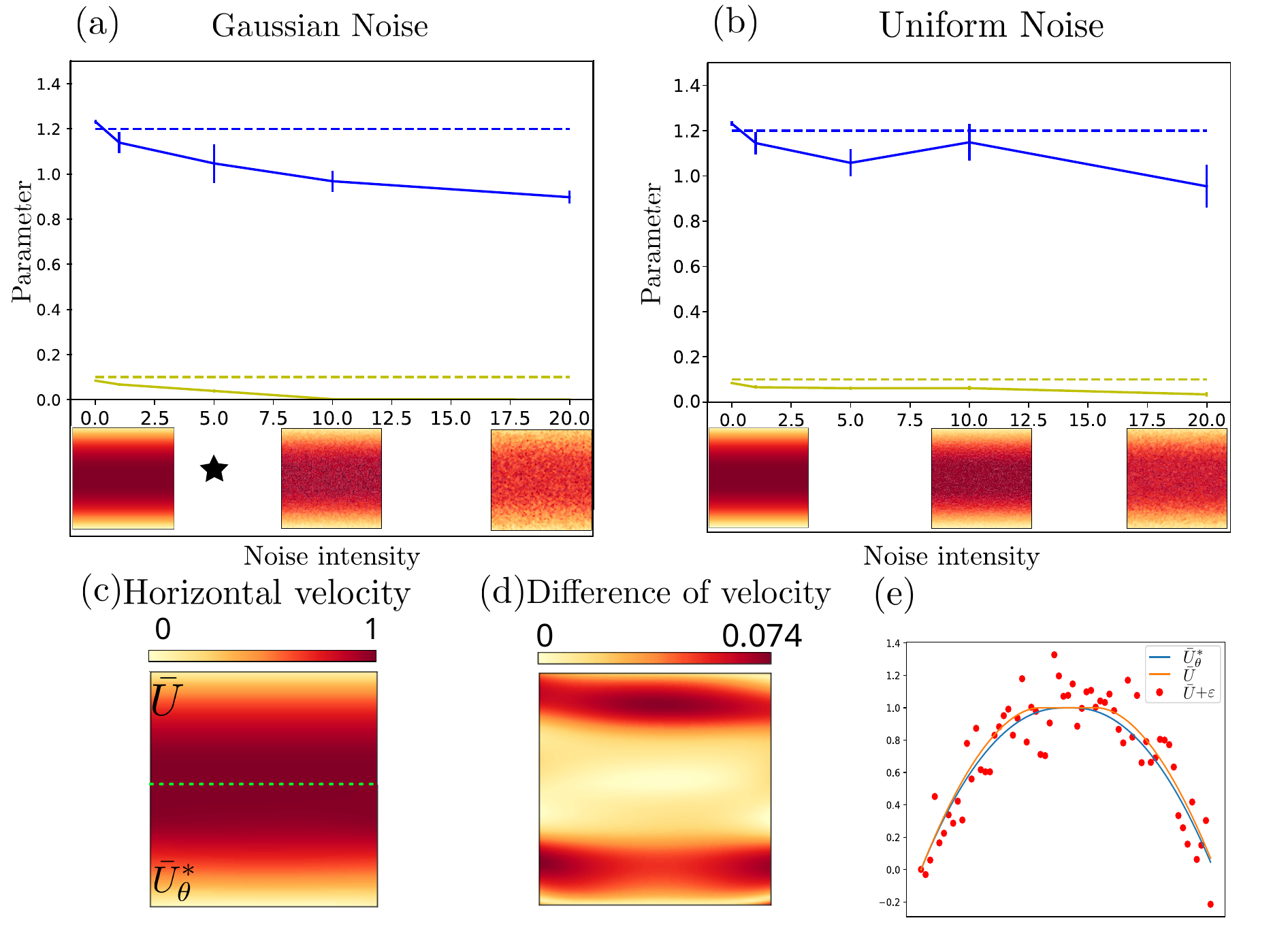}
    \caption{Rheology inference on noisy velocity data.
    (a,b) Mean values of the infered Herschel-Bulkley parameters in a uniform channel, over a series of 20 inference runs, for (a) Gaussian and (b) uniform noise of various amplitudes. 
    The error bars indicate the standard error of the mean. 
    Snapshots below the figures show the noisy velocity fields.
    (c) Velocity field rectonstruction in the case of a $5\%$ Gaussian noise (star symbol in (a)): (top) original velocity field, before addition of noise, (bottom) infered velocity field at the output of the neural network.
    (d) Absolute difference between the original and infered velocity fields.
    (e) Velocity profile along the vertical direction at the center of the channel in the case of a $20\%$ Gaussian noise: original velocity profile, sample points after addition of noise and infered velocity profile.}
    \label{fig:noise}
\end{figure}

An important challenge for the inference algorithm is to be able to perform on noisy experimental data.
To explore this aspect, we add artificial noise to the velocity data provided to our algorithm.
As the nature of the noise that needs to be removed from experimental data is generally unknown, we consider two different types of artificial noise, namely a gaussian and a uniform noise.
We perform series of inference runs with various noise amplitudes, corresponding to various standard deviation values for the gaussian noise and various limit values for the uniform noise, both normalized by the mean flow velocity.

Figure \ref{fig:noise} shows the inference results in the case of Herschel-Bulkley flow in a uniform channel.
For both types of noise, the inference accuracy decreases with the noise amplitude (figure \ref{fig:noise}a,b).
However, even with a 10\% noise amplitude, the inferred parameters remain reasonably close to their true values, especially in the case of uniform noise.
At the end of the inference process, the interpolated velocity at the output of the neural network is entirely denoised and matches the original velocity data (\ref{fig:noise}(c,d)).
This illustrates that the physical loss $\mathcal{L}_{phy}$ acts as a physics-based regularization function during the velocity interpolation.

Interestingly, the presence of noise does not increase the inference uncertainty (quantified by the error bars in figure \ref{fig:noise}a,b), but rather introduces a consistent bias towards a Newtonian rheology. 
This is further illustrated in figure \ref{fig:noise}e, which shows (i) the original velocity profile accross the channel, (ii) the sampling points accross the same cross-section after addition of a 20\% Gaussian noise, and (iii) the infered velocity at output of the neural network.
The reconstructed velocity indeed lacks the non-newtonian features observed in the original data, especially the central plug typical of viscoplastic channel flows.
However, it is also clear from the figure that noise amplitude is large compared to the velocity difference between the Newtonian and non-Newtonian flows; capturing such non-Newtonian effects in a very noisy environment thus appears to be challenging for the method.

\section{Biased sampling in different geometries}
\label{sec:dfferent_geo}

\begin{figure}[!t]
    \centering
        \includegraphics[width=\textwidth]{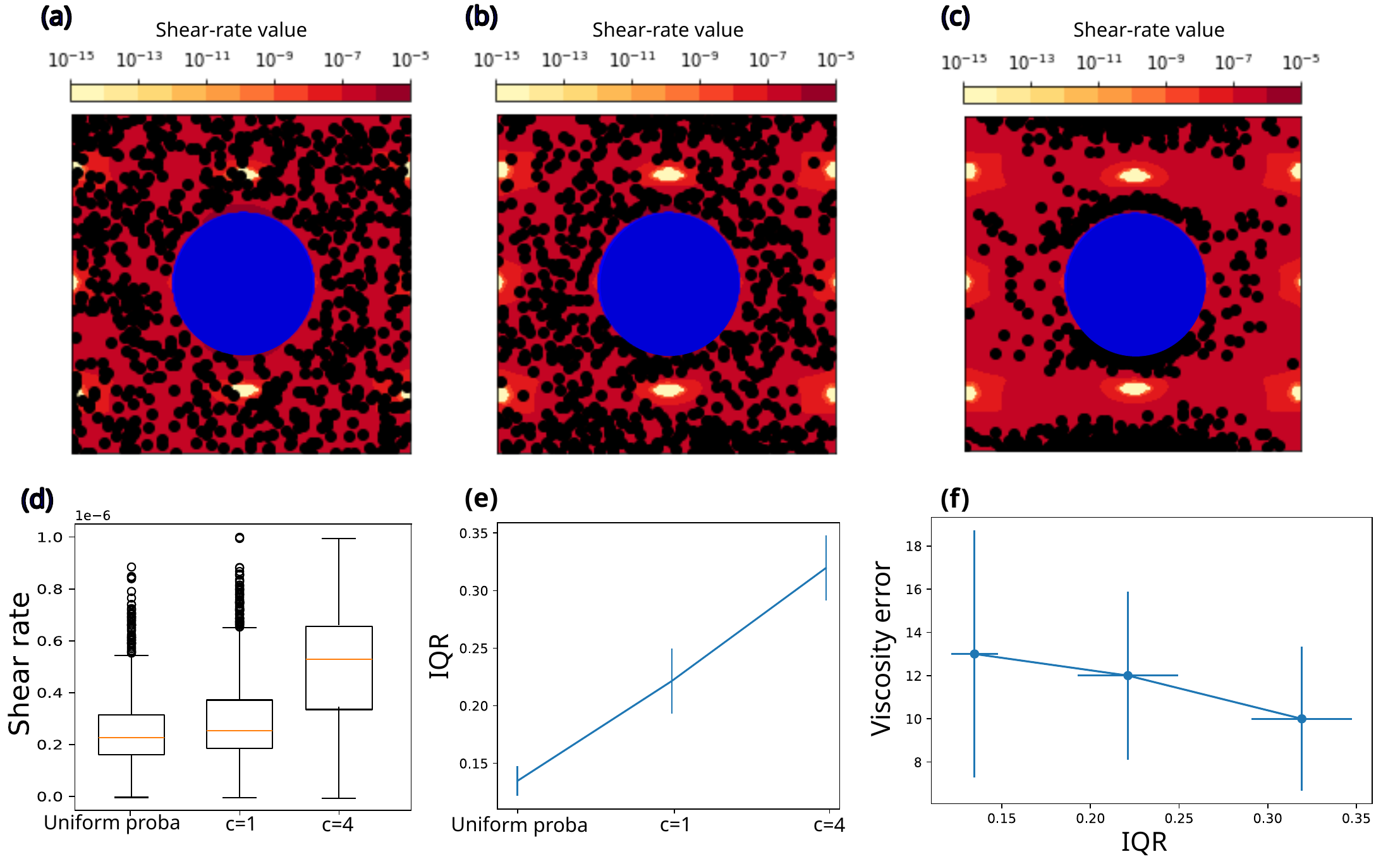}
    \caption{Illustration of the shear-biased sampling in the case of the flow past an obstacle.
    (a-c) Sampling point position for different values of the bias parameter $c$: (a) uniform sampling ($c=0$), (b) $c=1$ and (c) $c=4$.
    (d) Box plots showing the statistical shear-rate distribution in each case. 
    In each plot the horizontal lines correspond to the first and third quartiles, used to compute the inter-quartile range (IQR).
    (e) IQR as a function of the bias parameter $c$.
    (f) Inference error (\ref{eqn:error}) as a function of the bias parameter $c$.
    }
    \label{fig:samplingexamples}
\end{figure}

Accurately infering viscosity laws requires to sample data accross a broad range of shear rates.
However the shear rate distribution may vary from one geometry to another, implying a geometry-bias in the rheology inference.
Here, we propose to control this bias by performing a spatially non-uniform sampling, taking into account the shear-rate distribution in the available data.
We noticed that in the considered configurations, the shear rate tends to be naturally distributed towards lower values.
A first simple approach in order to alter the shear-rate distribution consists in biasing it towards higher values. 
Given access to the velocity field, we first compute the corresponding shear rate field. 
By binning this field, we determine the distribution of shear rate values across the bins, which provides a distribution denoted by $\Gamma(X)$. 
We then construct a new sampling probability distribution based on this information:

\begin{equation}
    P_s(X)=\frac{\Gamma(X)^c}{\sum_i[\Gamma(X_i)^c]},
    \label{eqn:proba}
\end{equation}
with $c$ a user-defined constant controlling the bias magnitude and $(X_i)$ being the points of the field.
Equation (\ref{eqn:proba}) provides a simple way to promote sampling in high-shear regions.

Before the training process, we sample velocity data according to the probability distribution $P_s$ (\ref{eqn:proba}).
Figure \ref{fig:samplingexamples} illustrates this biased sampling in the case of the flow past an obstacle.
As the parameter $c$ increases, the spatial distribution of the sampling points gets spatially non-uniform (figure \ref{fig:samplingexamples}a-c) and the statistical distribution of shear-rate values gets broader (figure \ref{fig:samplingexamples}d).
We quantify this aspect using the inter-quartile range (IQR) of the shear-rate distribution.
The IQR provides a robust measure of the diversity of information within the distribution, as it is insensitive to extreme values.
The IQR increases as the sampling gets more biased (figure \ref{fig:samplingexamples}e), which appears to significantly improve the inference accuracy.
We interpret this as the beneficial effect of improving the diversity of shear-rate values over which the viscosity law can be learned.

\begin{figure}[!t]
    \centering
    \includegraphics[width=\textwidth]{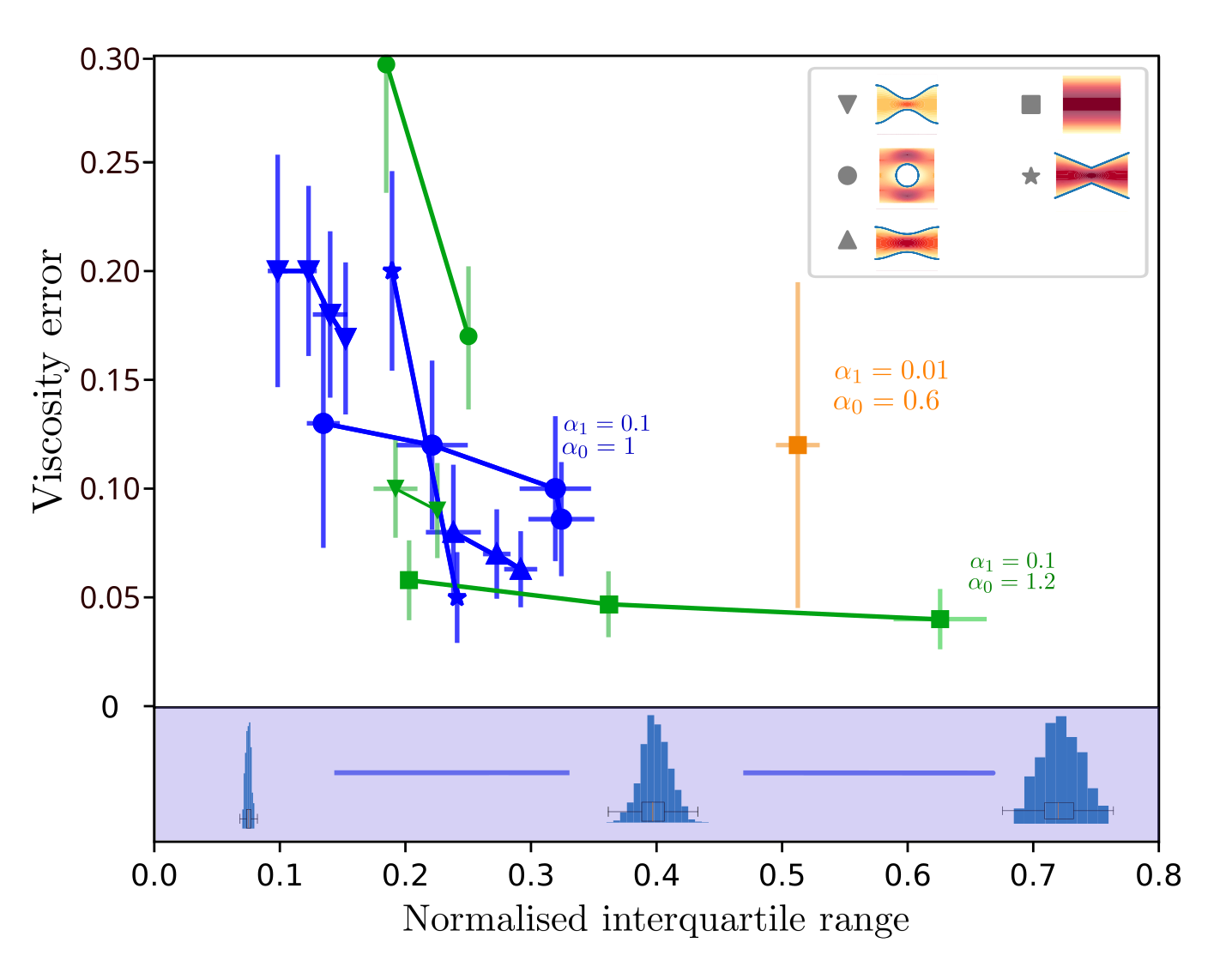}
    \caption{Effect of the inter-quartile range (IRQ) of the shear-rate distribution on the inference accuracy, for different configurations and different rheological parameters.
    Solid lines connect points with identical geometry and rheology, but different IQR achieved through biased sampling. 
    Each point is averaged over 20 inference runs performed with a fixed value of the bias parameter $c$.
    Horizontal error bars indicate the standard error on the mean IQR value, as data points are re-sampled for each inference run.
    Vertical error bars represent the error propagated from the standard error of the mean of the rheological parameters.
    }
    \label{fig:IQRvslogmse}
\end{figure}

We apply the same sampling strategy to various cases in different geometries and summarize the results in figure \ref{fig:IQRvslogmse}.
Even though variations are observed from one geometry to another, this confirms that increasing the IQR generally improves the inference quality. 
This also confirms that the inference process can be applied to a variety of geometries, on a broad range of rheological parameters values. 
The bias parameter $c$ thus provides a useful way to tune the sampling process depending on the shear-rate distribution within the available data, in order to improve the inference quality.

\section{Automatic rheology model selection}

So far we have illustrated the capacity of our PINN algorithm to fit rheological parameters based on velocity data, provided that the correct rheological model is known. 
We now adress the more challenging problem of rheological model selection: is it possible to select the most efficient model among a list of model candidates, while fitting the corresponding parameters? 
We consider a set of $L$ candidate viscosity laws , $\eta^*_1,\eta^*_2,..., \eta^*_L$, each with an associated pressure scaling parameter $\kappa_i$ and their corresponding number of viscosity parameters $n_1,...n_L$ (i.e, size of the set $\alpha$).
Each viscosity model is associated to a physical loss, 
\begin{equation}
    \mathcal{L}_{phy,i}=\lVert -\kappa_i\nabla P_\theta^*+ \nabla\cdot (2 \eta_i^* S^*_\theta) \rVert.
\end{equation}
We build the total physical loss using a function analogous to the Akaike Information Criterion (AIC) used in Bayesian inference \cite{murphy2012machine} or in some sparse regression algorithms using a $l_0$-norm regularization (\cite{weston2003use,pang2018efficient,chen2021physics}),
\begin{equation}
    \mathcal{L}_{phy,tot}=\sum_i \beta_i( \mathcal{L}_{phy,i}+ \lVert \eta_i \rVert_0),
    \label{eqn:Lphy_tot}
\end{equation}
with $\beta_i \in [0,1]$ an activation function following
\begin{equation}
    \beta_i(b_i)=\dfrac{\tanh(r b_i)+1}{2},
\end{equation}
and $b_i$ are model activation parameters that need to be learned during the training process.
The parameter $r$ sets the transition slope of the activation function; in the following we set $r = 10$.
The total physical loss (\ref{eqn:Lphy_tot}) depends on the number of physical parameters $\lVert \eta_i \rVert_0=n_i$ involved in each model, avoiding a bias towards the selection of the model involving the largest number of fit parameters.
We also need our algorithm to promote sparsity of the resulting rheological model by promoting the selection of a single rheological model during the learning process, i.e. setting one $\beta_i$ to $1$ and the others to $0$.
We thus add a sparsity loss analoguous to sparsity losses mentionned in \cite{narasimhan2018learning, cotter2019optimization}
\begin{equation}
  \mathcal{L}_\beta=2 \max_i(n_i) \lvert 1-\sum_i \beta_i \rvert. 
  \label{eqn:sparse_loss}
\end{equation}
The factor $2 \max_i(n_i)$ is included to prevent the scenario where $\beta_i = 0$ for all $i$.
Indeed, if $\beta_i = 0$ for all $i$ then $\mathcal{L}_\beta = 2 \max_i(n_i)$, which is greater than $n_i$ for all $i$. 
Therefore,
\begin{equation}
(\mathcal{L}_\beta+\mathcal{L}_{phy,tot})_{\beta=\{0,...,0\}}>(\mathcal{L}_\beta+\mathcal{L}_{phy,tot})_{\beta=\{0,...,1,...,0\}},
\end{equation}
which prevents the existence of a local minimum when all $\beta_i = 0$.
The sparsity loss (\ref{eqn:sparse_loss}) also prevents the case where $\beta_i = 1/L$ for all models. 

Indeed, 
\begin{align}
    (\mathcal{L}_{phy,tot}+\mathcal{L}_\beta)_{\beta=\{\frac{1}{L},...,\frac{1}{L}\}} &= \frac{1}{N} (\sum^L n_i+ \mathcal{L}_{phy,i}) \\
    &> \min_i (n_i+ \mathcal{L}_{phy,i}),
\end{align}
i.e.
\begin{equation}
    (\mathcal{L}_{phy,tot}+\mathcal{L}_\beta)_{\beta=\{\frac{1}{L},...,\frac{1}{L}\}} > (\mathcal{L}_{phy,tot}+\mathcal{L}_\beta)_{\beta=\{0,...,1,...,0\}}.
\end{equation}
Taken togher, the total model selection loss reads 
\begin{equation}
    \mathcal{L}_{MS}=\mathcal{L}_{data}+w\mathcal{L}_{phy,tot}+\mathcal{L}_p+\mathcal{L}_\beta.
    \label{eqn:total_loss_ms}
\end{equation}

We split the learning process into two phases.
First, we perform $3 \times 10^4$ epochs using the model selection loss (\ref{eqn:total_loss_ms}).
Then, we stop the model selection by choosing the model with the hightest value of $\beta_i$.
The associated activation function is set to 1 and the others to 0.
We finally perform additional epochs using the single-model loss (\ref{eqn:total_loss}) to fit the rheological parameters.

\begin{figure}[!t]
    \centering
        \includegraphics[width=\textwidth]{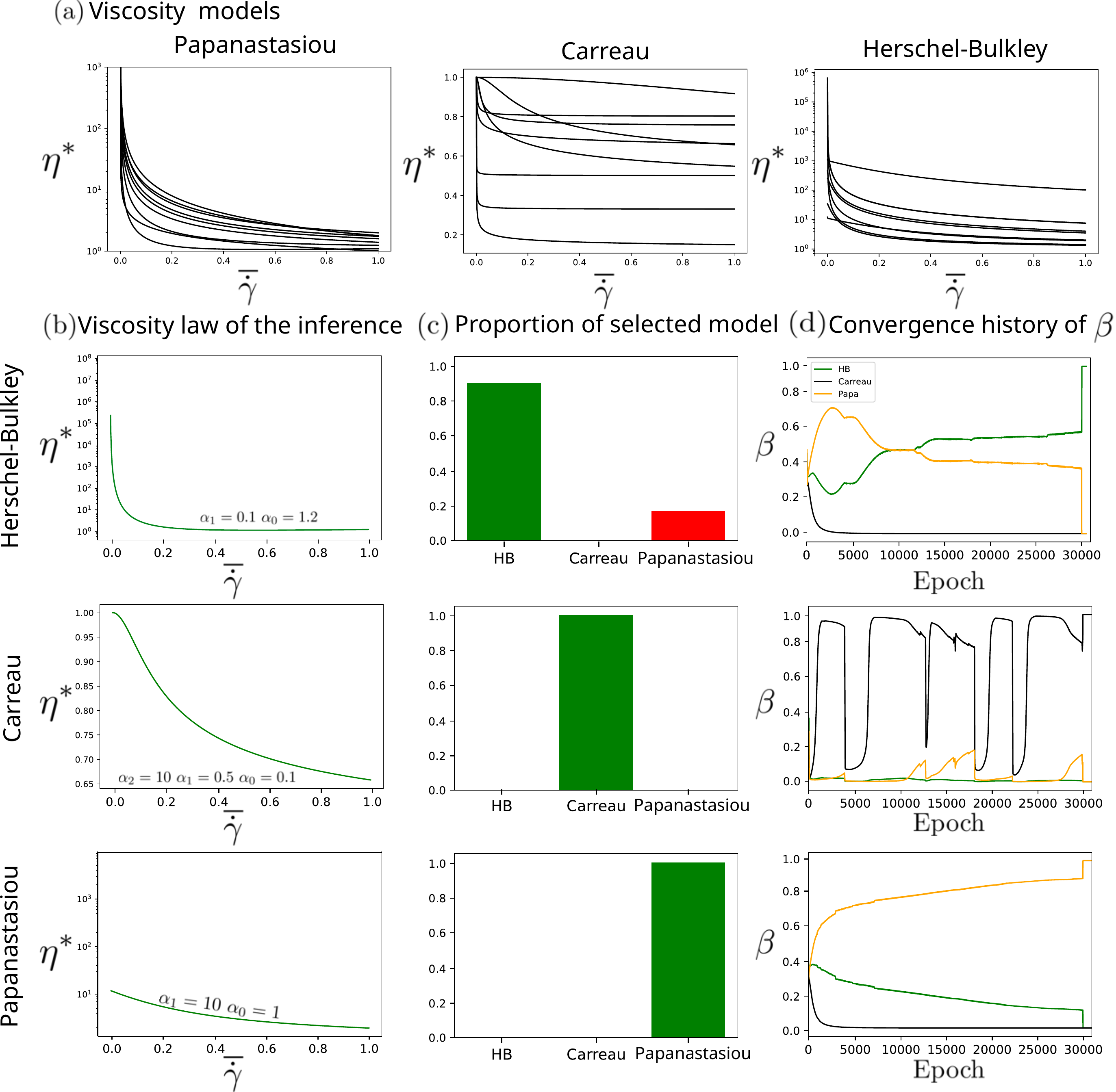}
    \caption{Rheology model selection.
    (a) Overview of the various viscosity functions $\eta^* (\dot{\gamma})$ possibly generated by the Herschel-Bulkley, Carreau and Papanastasiou models as their rheological parameters are varied. 
    For each model, the green line indicates the actual case provided to the algorithm to test our model selection approach. With $\bar{\dot{\gamma}}=\frac{\dot{\gamma}-\min(\dot{\gamma})}{\max(\dot{\gamma})-\min(\dot{\gamma})}$
    (b) Set of parameters with their associated viscosity provided to the algorithm to test our model selection approach.
    (c) Proportion of selected model over 30 tests when passing a simulation with the same set of parameters as in the column first column.
    (d) 3 examples of the convergence history of the 3 weights $\beta_i$}
    
    \label{fig:modelsel}
\end{figure}

We illustrate the performance of this approach in figure \ref{fig:modelsel}.
We provide the algorithm with three model candidates: Herschel-Bulkley, Carreau, and Papanastasiou.
It is worth noting that the Papanastasiou model is a regularized version of the Herschel-Bulkley model for $n=1$, meaning there exists a set of parameters $\alpha$ for which the Papanastasiou and Herschel-Bulkley viscosity laws are equivalent. 
To facilitate effective model selection, we constrain the Papanastasiou viscosity parameters $\alpha_0$ and $\alpha_1$ to be greater than 1 by adding a term $\max(0,1-\alpha_i)$ in the loss $\mathcal{L}_p$.
We select three examples in which the algorithm is applied to velocity data generated with one of the three models. 
In each example, we perform a total of 30 inference runs and show the resulting proportion of selected models in figure \ref{fig:modelsel}c.
In all cases, the algorithm selects the correct rheological model, except for a few runs in which the Panastasiou and Herschel-Bulkley models are confused, especially when the true model is the Herschel-Bulkley model. 
This can be attributed to the similarity between both models for certain parameter sets (see figure \ref{fig:modelsel}).
As shown in figure \ref{fig:modelsel}d, the dynamics of the activation functions during training varies from one model to the other, but in all cases the correct model is rapidly promoted compared to other model candidates.
Overall, the proposed model selection method is effective and easely extendable to a larger number of candidate models.

\section{Discusion}

Since their introduction, physics-informed neural networks (PINNs) \cite{raissi2019physics} have seen rapid development and adoption across a range of physical and engineering disciplines. 
Their strength particularly lies in solving inverse problems—inferring physical properties or governing models from data—by leveraging both observational data and physical laws to constrain solutions. 
This makes them especially well-suited for tasks such as rheological model inference, where the underlying constitutive behavior is often complex or partially unknown.

Recent studies have begun to explore the application of PINNs to rheological problems, either by using classical rheometry data \cite{mahmoudabadbozchelou2021rheology, mahmoudabadbozchelou22} or, more recently, by learning from velocity fields \cite{thakur2024viscoelasticnet, zhai2023deep}. 
In this work, we demonstrate that even relatively simple PINN architectures—consisting of a single fully connected neural network—are capable of inferring viscoplastic properties directly from noisy velocity data. 
Notably, our approach does not require prior knowledge of boundary conditions, making it particularly promising for analyzing experimental data, where stress boundary conditions are typically unknown and the no-slip assumption at solid boundaries may not hold \cite{cheddadi11, golovkova20, tlili20, tlili_microfluidic_2022}.

The methodology we propose is designed to be readily extensible to a wide range of viscoplastic fluid types. 
In addition, we introduce a practical strategy for model selection during the inference process, enabling automatic identification of the most suitable constitutive law from a set of candidate models. 
This adds a significant layer of flexibility and robustness to the framework, potentially easing its adoption in experimental and industrial settings.

Several directions for future research naturally emerge from this work. 
A primary next step is the application of our approach to experimental velocity fields. This would involve benchmarking the method on systems with well-characterized rheology, followed by its deployment on materials for which traditional characterization techniques are limited or infeasible. 
Furthermore, the framework can be extended to accommodate more complex rheological behaviors, such as viscoelastic or viscoplastoelastic responses. 
One particularly compelling objective is to generalize the model selection capability to span these broader rheological classes, allowing the PINN to autonomously determine the most appropriate constitutive description for a given dataset.
Lastly, there remains a need for deeper theoretical understanding of PINN training dynamics, particularly in the context of inverse rheological problems. 
Issues such as convergence rates, robustness to noise, and identifiability of model parameters warrant further investigation. 
In the longer term, we envision that approaches such as the one developed here will play a central role in advancing computational rheology and in addressing interdisciplinary challenges, notably in the study of biological and geophysical materials, where conventional modeling approaches often fall short.

\section*{Acknowledgements}

We thank Christophe Eloy, Hélène Delanoë-Ayari, Yoël Forterre, Philippe Roudot and Andonis Gerardos for useful discussion.
The project leading to this publication has received funding from the ‘‘Investissements d’Avenir’’ French Government program managed by the French National Research Agency (ANR-16-CONV-0001) and from ‘‘Excellence Initiative of Aix-Marseille University – A*MIDEX’’.
We also acknowledge the French National Research Agency, under grant ANR-24-CE45-1959.
The Centre de Calcul Intensif d'Aix-Marseille is acknowledged for granting access to its high performance computing resources.

\section*{Declaration of generative AI in the writing process}

The authors used ChatGPT in order to improve the writing quality of their manuscript. 
After using this tool/service, the authors reviewed and edited the content as needed and take full responsibility for the content of the publication.

\bibliographystyle{elsarticle-num} 
\bibliography{cas-refs2}

\begin{thebibliography}{10}
\expandafter\ifx\csname url\endcsname\relax
  \def\url#1{\texttt{#1}}\fi
\expandafter\ifx\csname urlprefix\endcsname\relax\def\urlprefix{URL }\fi
\expandafter\ifx\csname href\endcsname\relax
  \def\href#1#2{#2} \def\path#1{#1}\fi

\bibitem{lenne22}
P.-F. Lenne, V.~Trivedi, Sculpting tissues by phase transitions, Nature
  Communications 13~(1) (2022) 664.

\bibitem{davaille18}
A.~Davaille, P.~Carrez, P.~Cordier, Fat plumes may reflect the complex rheology
  of the lower mantle, Geophysical Research Letters 45~(3) (2018) 1349--1354.

\bibitem{tommasi2000viscoplastic}
A.~Tommasi, D.~Mainprice, G.~Canova, Y.~Chastel, Viscoplastic self-consistent
  and equilibrium-based modeling of olivine lattice preferred orientations:
  Implications for the upper mantle seismic anisotropy, Journal of Geophysical
  Research: Solid Earth 105~(B4) (2000) 7893--7908.

\bibitem{jop06}
P.~Jop, Y.~Forterre, O.~Pouliquen, A constitutive law for dense granular flows,
  Nature 441~(7094) (2006) 727--730.

\bibitem{guazzelli18}
{\'E}.~Guazzelli, O.~Pouliquen, Rheology of dense granular suspensions, Journal
  of Fluid Mechanics 852 (2018) P1.

\bibitem{etcheverry23}
B.~Etcheverry, Y.~Forterre, B.~Metzger, Capillary-stress controlled rheometer
  reveals the dual rheology of shear-thickening suspensions, Physical Review X
  13~(1) (2023) 011024.

\bibitem{Brunton_2016}
S.~L. Brunton, S.~L. Brunton, J.~L. Proctor, J.~L. Proctor, J.~N. Kutz, J.~N.
  Kutz, Discovering governing equations from data by sparse identification of
  nonlinear dynamical systems, Proceedings of the National Academy of Sciences
  of the United States of America (2016).
\newblock \href {https://doi.org/10.1073/pnas.1517384113}
  {\path{doi:10.1073/pnas.1517384113}}.

\bibitem{bongard2007automated}
J.~Bongard, H.~Lipson, Automated reverse engineering of nonlinear dynamical
  systems, Proceedings of the National Academy of Sciences 104~(24) (2007)
  9943--9948.

\bibitem{rudy2017data}
S.~H. Rudy, S.~L. Brunton, J.~L. Proctor, J.~N. Kutz, Data-driven discovery of
  partial differential equations, Science advances 3~(4) (2017) e1602614.

\bibitem{kaheman2020sindy}
K.~Kaheman, J.~N. Kutz, S.~L. Brunton, Sindy-pi: a robust algorithm for
  parallel implicit sparse identification of nonlinear dynamics, Proceedings of
  the Royal Society A 476~(2242) (2020) 20200279.

\bibitem{chen2021physics}
Z.~Chen, Y.~Liu, H.~Sun, Physics-informed learning of governing equations from
  scarce data, Nature communications 12~(1) (2021) 6136.

\bibitem{colen2021machine}
J.~Colen, M.~Han, R.~Zhang, S.~A. Redford, L.~M. Lemma, L.~Morgan, P.~V.
  Ruijgrok, R.~Adkins, Z.~Bryant, Z.~Dogic, et~al., Machine learning
  active-nematic hydrodynamics, Proceedings of the National Academy of Sciences
  118~(10) (2021) e2016708118.

\bibitem{christopher2018parameter}
J.~D. Christopher, N.~T. Wimer, C.~Lapointe, T.~R. Hayden, I.~Grooms, G.~B.
  Rieker, P.~E. Hamlington, Parameter estimation for complex thermal-fluid
  flows using approximate bayesian computation, Physical Review Fluids 3~(10)
  (2018) 104602.

\bibitem{kontogiannis2024learning}
A.~Kontogiannis, R.~Hodgkinson, E.~L. Manchester, Learning rheological
  parameters of non-newtonian fluids from velocimetry data, arXiv preprint
  arXiv:2408.02604 (2024).

\bibitem{fan2020solving}
T.~Fan, K.~Xu, J.~Pathak, E.~Darve, Solving inverse problems in steady-state
  navier-stokes equations using deep neural networks, arXiv preprint
  arXiv:2008.13074 (2020).

\bibitem{xu21}
K.~Xu, A.~M. Tartakovsky, J.~Burghardt, E.~Darve, Learning viscoelasticity
  models from indirect data using deep neural networks, Computer Methods in
  Applied Mechanics and Engineering 387 (2021) 114124.

\bibitem{brunton2020machine}
S.~L. Brunton, B.~R. Noack, P.~Koumoutsakos, Machine learning for fluid
  mechanics, Annual review of fluid mechanics 52~(1) (2020) 477--508.

\bibitem{raissi2019physics}
M.~Raissi, P.~Perdikaris, G.~E. Karniadakis, Physics-informed neural networks:
  A deep learning framework for solving forward and inverse problems involving
  nonlinear partial differential equations, Journal of Computational physics
  378 (2019) 686--707.

\bibitem{lagaris1998artificial}
I.~E. Lagaris, A.~Likas, D.~I. Fotiadis, Artificial neural networks for solving
  ordinary and partial differential equations, IEEE transactions on neural
  networks 9~(5) (1998) 987--1000.

\bibitem{baydin18}
A.~G. Baydin, B.~A. Pearlmutter, A.~A. Radul, J.~M. Siskind, Automatic
  differentiation in machine learning: a survey, Journal of machine learning
  research 18~(153) (2018) 1--43.

\bibitem{cai2021physics}
S.~Cai, Z.~Mao, Z.~Wang, M.~Yin, G.~E. Karniadakis, Physics-informed neural
  networks (pinns) for fluid mechanics: A review, Acta Mechanica Sinica 37~(12)
  (2021) 1727--1738.

\bibitem{raissi20}
M.~Raissi, A.~Yazdani, G.~E. Karniadakis, Hidden fluid mechanics: Learning
  velocity and pressure fields from flow visualizations, Science 367~(6481)
  (2020) 1026--1030.

\bibitem{lou2021physics}
Q.~Lou, X.~Meng, G.~E. Karniadakis, Physics-informed neural networks for
  solving forward and inverse flow problems via the boltzmann-bgk formulation,
  Journal of Computational Physics 447 (2021) 110676.

\bibitem{zhai2023deep}
R.~Zhai, D.~Yin, G.~Pang, A deep learning framework for solving forward and
  inverse problems of power-law fluids, Physics of Fluids 35~(9) (2023).

\bibitem{thakur2024viscoelasticnet}
S.~Thakur, M.~Raissi, A.~M. Ardekani, Viscoelasticnet: A physics informed
  neural network framework for stress discovery and model selection, Journal of
  Non-Newtonian Fluid Mechanics (2024) 105265.

\bibitem{mahmoudabadbozchelou2021rheology}
M.~Mahmoudabadbozchelou, S.~Jamali, Rheology-informed neural networks (rhinns)
  for forward and inverse metamodelling of complex fluids, Scientific reports
  11~(1) (2021) 12015.

\bibitem{mahmoudabadbozchelou22}
M.~Mahmoudabadbozchelou, K.~M. Kamani, S.~A. Rogers, S.~Jamali, Digital
  rheometer twins: Learning the hidden rheology of complex fluids through
  rheology-informed graph neural networks, Proceedings of the National Academy
  of Sciences 119~(20) (2022) e2202234119.

\bibitem{cheddadi11}
I.~Cheddadi, P.~Saramito, B.~Dollet, C.~Raufaste, F.~Graner, Understanding and
  predicting viscous, elastic, plastic flows, The European Physical Journal E
  34 (2011) 1--15.

\bibitem{golovkova20}
I.~Golovkova, L.~Montel, E.~Wandersman, T.~Bertrand, A.~M. Prevost, L.-L.
  Pontani, Depletion attraction impairs the plasticity of emulsions flowing in
  a constriction, Soft Matter 16~(13) (2020) 3294--3302.

\bibitem{tlili20}
S.~Tlili, M.~Durande, C.~Gay, B.~Ladoux, F.~Graner, H.~Delano{\"e}-Ayari,
  Migrating epithelial monolayer flows like a maxwell viscoelastic liquid,
  Physical Review Letters 125~(8) (2020) 088102.

\bibitem{tlili_microfluidic_2022}
S.~L. Tlili, F.~Graner, H.~Delanoë-Ayari, A microfluidic platform to
  investigate the role of mechanical constraints on tissue reorganization
  149~(20)  dev200774.
\newblock \href {https://doi.org/10.1242/dev.200774}
  {\path{doi:10.1242/dev.200774}}.

\bibitem{papanastasiou1987flows}
T.~C. Papanastasiou, Flows of materials with yield, Journal of rheology 31~(5)
  (1987) 385--404.

\bibitem{bonn2017yield}
D.~Bonn, M.~M. Denn, L.~Berthier, T.~Divoux, S.~Manneville, Yield stress
  materials in soft condensed matter, Reviews of Modern Physics 89~(3) (2017)
  035005.

\bibitem{kingma2014adam}
D.~P. Kingma, J.~Ba, Adam: A method for stochastic optimization, arXiv preprint
  arXiv:1412.6980 (2014).

\bibitem{lardygit}
\url{https://github.com/martinLARD/Pinn.git}.

\bibitem{gsell2021lattice}
S.~Gsell, U.~d'Ortona, J.~Favier, Lattice-boltzmann simulation of creeping
  generalized newtonian flows: theory and guidelines, Journal of Computational
  Physics 429 (2021) 109943.

\bibitem{gsell2021direct}
S.~Gsell, J.~Favier, Direct-forcing immersed-boundary method: a simple
  correction preventing boundary slip error, Journal of Computational Physics
  435 (2021) 110265.

\bibitem{chen2025global}
F.~Chen, Y.~Meng, K.~Li, C.~Yang, J.~Yang, Global physics-informed neural
  networks (gpinns): from local point-wise constraint to global nodal
  association, arXiv preprint arXiv:2503.06403 (2025).

\bibitem{murphy2012machine}
K.~P. Murphy, Machine learning: a probabilistic perspective, MIT press, 2012.

\bibitem{weston2003use}
J.~Weston, A.~Elisseeff, B.~Sch{\"o}lkopf, M.~Tipping, Use of the zero-norm
  with linear models and kernel methods, Journal of machine learning research
  3~(Mar) (2003) 1439--1461.

\bibitem{pang2018efficient}
T.~Pang, F.~Nie, J.~Han, X.~Li, Efficient feature selection via $l_{2,0}$-norm
  constrained sparse regression, IEEE Transactions on Knowledge and Data
  Engineering 31~(5) (2018) 880--893.

\bibitem{narasimhan2018learning}
H.~Narasimhan, Learning with complex loss functions and constraints, in:
  International Conference on Artificial Intelligence and Statistics, PMLR,
  2018, pp. 1646--1654.

\bibitem{cotter2019optimization}
A.~Cotter, H.~Jiang, M.~Gupta, S.~Wang, T.~Narayan, S.~You, K.~Sridharan,
  Optimization with non-differentiable constraints with applications to
  fairness, recall, churn, and other goals, Journal of Machine Learning
  Research 20~(172) (2019) 1--59.

\end{thebibliography}

\end{document}